\documentclass{article}

\usepackage{arxiv}

\usepackage[utf8]{inputenc} % allow utf-8 input
\usepackage[T1]{fontenc}    % use 8-bit T1 fonts
\usepackage{hyperref}       % hyperlinks
\usepackage{url}            % simple URL typesetting

\usepackage{graphicx}
\usepackage{natbib}
\usepackage{doi}

\usepackage{caption}

\title{Compendium Manager: a tool for coordination of workflow management instances for bulk data processing in Python}

\author{ Richard J. Abdill\thanks{Correspondence to \texttt{rabdill@uchicago.edu}.} \hspace{1mm} and \hspace{1mm}Ran Blekhman\\
	Section of Genetic Medicine\\
        Department of Medicine\\
	University of Chicago, Chicago, IL 60637
}
\date{}

\renewcommand{\undertitle}{}
\renewcommand{\shorttitle}{Compendium Manager}

\hypersetup{
pdftitle={Compendium Manager: a tool for coordination of workflow management instances for bulk data processing in Python},
pdfsubject={q-bio.OT},
pdfauthor={Richard J. Abdill, Ran Blekhman},
pdfkeywords={compendium, bioinformatics, microbiome, public data},
}

\begin{document}
\maketitle

\begin{abstract}
    Compendium Manager is a command-line tool written in Python to automate the provisioning, launch, and evaluation of bioinformatics pipelines. Although workflow management tools such as Snakemake and Nextflow enable users to automate the processing of samples within a single sequencing project, integrating many datasets in bulk requires launching and monitoring hundreds or thousands of pipelines. We present the Compendium Manager, a lightweight command-line tool to enable launching and monitoring analysis pipelines at scale. The tool can gauge progress through a list of projects, load results into a shared database, and record detailed processing metrics for later evaluation and reproducibility.

\end{abstract}

% keywords can be removed

\section{Introduction}
The rise of high-throughput DNA sequencing confronted biologists with ``big data'' problems that they had previously been able to ignore. Biologists had performed software-based analysis tasks for decades (Ouzounis 2012; Ouzounis and Valencia 2003), including work with DNA sequences (Pearson and Lipman 1988; Altschul et al. 1990), but this new world, in which a single study's DNA sequencing could generate terabytes of data, produced challenges impossible to address on consumer-grade workstations (Marx 2013).

In the 2010s, as more biologists added computational skills and the number of sequencing projects expanded dramatically (Katz et al. 2022), bioinformatics software became more commonly built (Mangul et al. 2019), used, and cited (Wren 2016). Genomics processing frequently requires complicated, multi-step protocols in which the output of one (or more) tools is used as the input to subsequent tools, with many intermediate files being passed between processes. In smaller projects, it's possible for a researcher to follow these steps (the ``pipeline'') from a terminal window, monitoring progress and launching each step when ready. But it quickly becomes impractical to manually usher hundreds or thousands of samples through these pipelines in a consistent way. This concern was addressed with automation tools called workflow managers or bioinformatic pipeline frameworks (Leipzig 2017): Tools such as Snakemake (Köster and Rahmann 2012) operate by following information from a text file describing each step of the pipeline, such as the commands that must be run, resources required for the step, and how the environment should be configured. Users define their pipeline, which may have one or dozens of steps (Pohl et al. 2024), apply it to any collection of samples, and let the management software iterate through all steps for all samples.

A common approach is to structure a pipeline to apply to a single project or sequencing run: tools such as DADA2 include experiment-specific error modeling (Callahan et al. 2016), for example. In ordinary circumstances, this is the usual way data would be processed anyway---however, researchers who seek to process many projects again encounter the problem of scale requiring an impractical level of human intervention.

Meta-analysis efforts, which are becoming increasingly common as genomic data accumulates, may involve integrating hundreds or thousands of independent sequencing projects, each requiring its own pipeline instance with project-specific parameters and error handling. Although workflow managers like Snakemake excel at automating steps within a pipeline, they lack built-in capabilities for coordinating multiple independent pipeline instances across different projects, monitoring their collective progress, and consolidating results into a unified database. This gap often leads researchers to rely on ad-hoc scripts, manual monitoring, and fragmented approaches to project management. This creates bottlenecks that limit the scale of integrative analyses and introduces inconsistencies in how datasets are processed. To address this gap, we developed the Compendium Manager, a lightweight yet robust orchestration layer that provides automated coordination of multiple workflow instances, centralized error handling, and integrated result management for large-scale bioinformatics meta-analyses.

Several tools are already available to address portions of this situation: The snk package can generate customizable command-line interfaces for conventional Snakemake pipelines (Wirth, Mutch, and Turnbull 2024), and Snakemaker uses large language models to parse command-line activity and generate pipeline code (Masera et al. 2025). It is possible to use Snakemake itself to process multiple projects independently, but when processing all projects could take weeks or months given resource constraints, having a single Snakemake process monitoring all of this work would be impractical and error-prone, particularly in a shared high-performance computing environment that likely imposes time restrictions.

Existing compendia solve this orchestration issue in several ways: the recount3 project, a dataset of 730,000 human and mouse RNA-seq samples (Wilks et al. 2021), also uses Snakemake to process individual projects. These individual project pipelines are coordinated using their ``Monorail'' system, which distributes work across nodes either in the cloud (e.g. Amazon Web Services servers) or on HPC clusters before a ``unify'' step collects data together. The ARCHS4 project also used cloud resources; it used processing tasks automated via Python and SQLite (Lachmann et al. 2018), as the Compendium Manager does. However, its suite of scripts does not include a user interface for performing tasks such as processing projects or evaluating results (Clarke, Ma’ayan, and Lachmann 2020). The curatedMetagenomicData package, which contains data (and manually curated metadata) for more than 20,000 shotgun samples (Pasolli et al. 2017), has documentation describing a pipeline in which each step is launched manually by a curator (Home, n.d.). The GMrepo project also used DADA2 but did not specify any management-level automation (Dai et al. 2022). The Single-cell Pediatric Cancer Atlas automated project-level processing and merging using Nextflow (Di Tommaso et al. 2017), but does not describe compendium-level automation (Hawkins et al. 2024).

\section{Compendium Manager overview}
Despite the utility of workflow management tools in a high-performance computing environment, there is no broadly accepted solution for ``automating the automation''---that is, coordinating many Snakemake pipelines operating independently of each other. Projects seeking to integrate many projects together may still use Snakemake, but the coordination of those pipelines is done manually (Valderrama et al. 2025). The Compendium Manager software addresses these issues by providing a customizable platform that performs the following tasks:
\begin{itemize}
    \item Recording and organizing metadata for a corpus of samples split into projects
    \item Provisioning project-specific pipeline environments
    \item Tracking each project's pipeline progress, errors encountered and the automated changes made to address them
    \item Providing a command-line interface to monitor the status of computational jobs and compendium progress, plus methods for processing new projects
    \item Storing project results in a database, providing a streamlined source of data for generating compendium files and integrated results
\end{itemize}

The current codebase and example Snakemake workflow reflects its use in the Human Microbiome Compendium project for processing 16S rRNA gene amplicon sequencing datasets (Abdill et al. 2025), but each of the Compendium Manager's tasks---starting new pipelines, monitoring progress, responding to errors, evaluating results, determining required re-processing steps, and so on---can be expanded to an arbitrary level of complexity for other types of data. In general, the Compendium Manager's approach is straightforward: Given the results of a BioProject search, the software builds a database of projects and their associated samples. Then, either automatically or at the prompting of a user, the tool will generate the required files for a given project, launch a Snakemake pipeline to process its files. If the final step in the Snakemake process is configured to report progress back to the Compendium Manager software, the tool will then evaluate the results, record them in the database if appropriate, archive or delete the remaining files, then launch processing for the next project (\textbf{Figure 1}).

\begin{figure}
	\centering
	\includegraphics[scale=0.5]{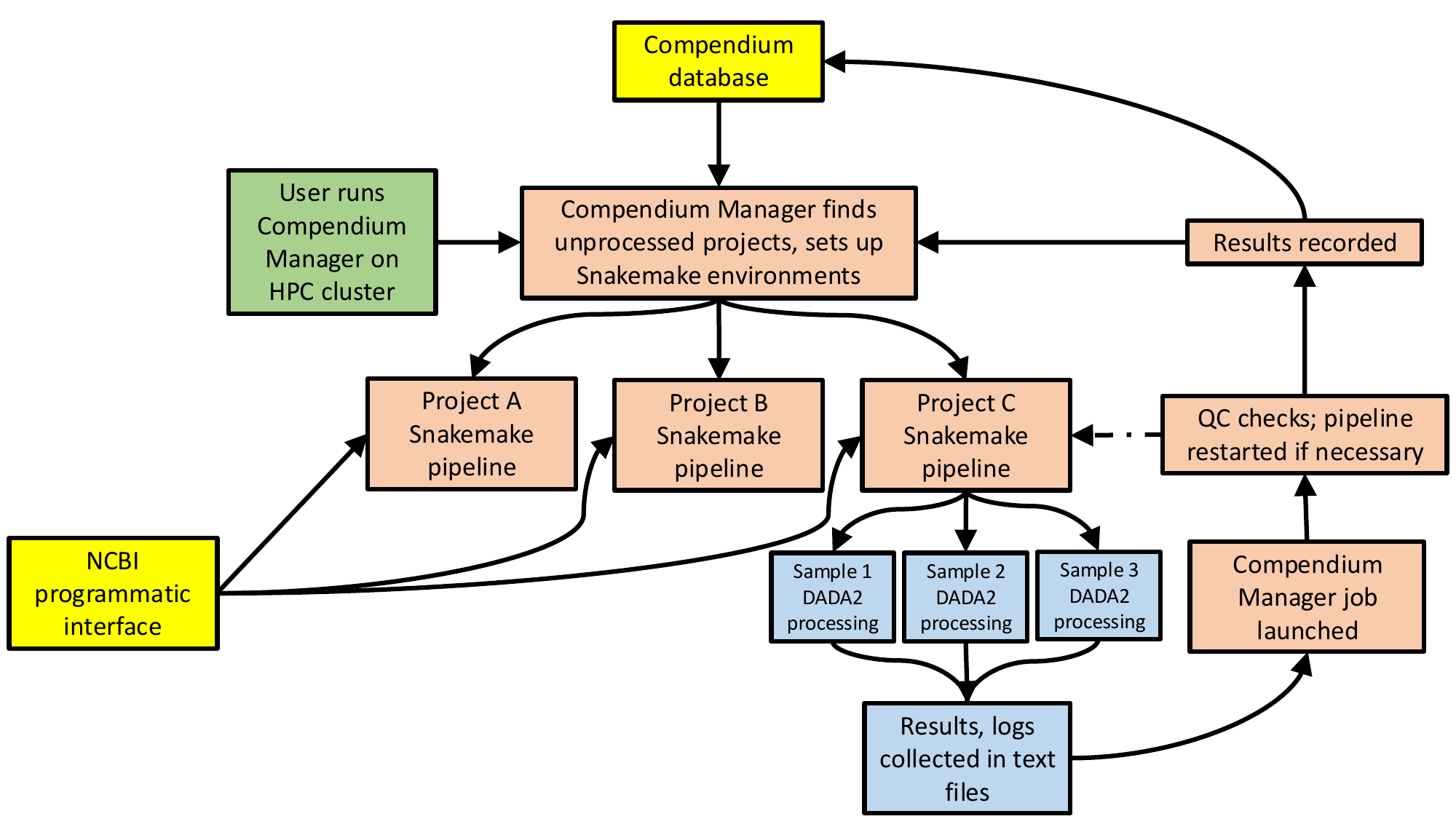}
    \captionsetup{format=hang}
	\caption{Workflow illustration. A flow chart indicating the processes followed at the prompting of a user for an example amplicon sequencing workflow. The green square indicates user interaction. Yellow squares are data sources such as databases or APIs. The orange boxes indicate steps launched and monitored by the Compendium Manager tool. Blue boxes indicate steps that are automated by Snakemake.}
	\label{fig:fig1}
\end{figure}

Though implementation details will differ between projects and data types, we are optimistic that our general approach is a practical solution to this task:
\begin{enumerate}
\item The user performs a search of the BioSample database (Barrett et al. 2012) via its web interface, defining the corpus of samples to be considered for inclusion in the compendium. These results are then exported as an XML document.
\item The Compendium Manager's ``xml'' command is used to parse this XML file and load sample metadata into its internal database: information such as library strategy, deposition date and project ID are recorded in a ``samples'' table in the database, and key/value pairs of free-text metadata are saved in a ``tags'' table.
\item At this point, all samples have a sample ID, but sequencing data is associated with runs. The package's ``runs'' command is used to query NCBI resources to determine the ID of each run associated with each sample.
\item Now, a user may use the SQLite database to filter out projects and samples that should not be processed: projects below a minimum sample count, for example, or samples reporting an uncommon sequencing platform.
\item Once prompted by the user to process a given BioProject, the package initializes Snakemake by downloading the pipeline definition files and generating a list of sequencing runs to be downloaded. The final step here is to launch a new compute job for the Snakemake program, which then processes the project using its own automation. The example workflow proceeds in this way:
    \begin{enumerate}
    \item Sample files are downloaded using the SRA Toolkit and converted into FASTQ format.
    \item Reads are filtered and trimmed to meet quality control thresholds.
    \item DADA2 is used to model sequencing errors observed for forward reads and, separately if applicable, reverse reads.
    \item These error models are incorporated into the DADA2 process for merging paired-end reads (if necessary) and filtering out reads that are chimeric or too short for taxonomic inference.
    \item A table of read counts is generated describing each sample in the project and the number of reads observed for each amplicon sequence variant (ASV).
    \item A summary file is generated describing the effect of each quality control step.
    \end{enumerate}
\item The final step in the Snakemake pipeline is to submit a new computed job that launches the Compendium Manager software with the ``autoforward'' command (see ``Usage and design'' below). This command parses the summary file to evaluate the results of the quantification.
    \begin{enumerate}
    \item If all user-defined thresholds are cleared for factors such as reads passing quality filters, then the software opens the results file, loads the ASV-level read counts into the database, and archives the intermediate files generated during processing of the project.
    \item If a threshold is violated, the project is re-processed using a pre-defined response to the observed violation. If the proportion of reads that were successfully merged with their mate is too low, for example, the project will be re-processed as single-end data.
    \item If there were quality violations and no response is available, the files are cleaned up as the project is discarded. Violations are also logged to the database to enable the user to return to discarded projects if a new approach is later added to the pipeline.
    \end{enumerate}
\item After a project's pipeline is completed and its results are successfully evaluated (resulting in either the project's results being recorded or discarded), the Compendium Manager software queries its database to find another project that still requires processing. If the user-defined limit to concurrent projects is not met, the Compendium Manager submits this new project for processing, and the protocol restarts at step 5 above.
\end{enumerate}

\section{Usage and design}
The command-line interface for the Compendium Manager organizes operations into two categories: ``compendium'' commands, which address status updates and metadata processing steps. The other category, ``project'' commands, deal with the processing of individual BioProjects.

\subsection{Compendium-level operations}

\begin{itemize}
\item \texttt{summary} -- Summarizes the content of the compendium.
\item \texttt{status} -- Summarizes the status of any projects with steps remaining in their processing pipeline.
\item \texttt{xml} -- Parse exported BioSample search results and load sample data into the database.
\item \texttt{runs} -- Queries the compendium database for samples that have an SRS (sample) number, but not an SRR (run) number. This list is then sent to the NCBI eUtils API to retrieve the runs. This is required for downloading the raw data.
\item \texttt{asvs} -- Runs a heuristic process for inferring which hypervariable regions were targeted in an amplicon sequencing project.
\item \texttt{forward} -- Interactive process for evaluating all currently pending projects.
\item \texttt{autoforward} -- Similar to the ``forward'' command, but automatically approves actions that need to be taken. If projects are completed, the application will then search for new projects to start. Unlike ``forward,'' this launches new projects as others are completed.
\end{itemize}

\subsection{Project-level operations}
\begin{itemize}
\item \texttt{runit} -- Initializes the processing pipeline for a single project and starts the pipeline. This is generally used to start the pipeline for the first time, because it creates a new directory for the project and pulls in all the necessary pipeline code. This will retrieve and process the actual FASTQ files.
\item \texttt{status} -- Retrieves the pipeline progress of a single project and prints a report to the terminal.
\item \texttt{eval} -- Evaluate the results of a single study. If it's completed the pipeline, it will evaluate the results and prompt the user to confirm that the project should either be saved and finalized, OR should be re-run with different parameters.
\item \texttt{again} -- Submits a new slurm job to restart the Snakemake pipeline for a single project. Used mostly for situations in which a project stalled for reasons that have been remediated---a missing sample, for example, or a timeout.
\item \texttt{discard} -- Throws out any computational results from a single project and records in the database that the project should not be re-attempted. Also prompts the user for a brief explanation of why it should be skipped.
\end{itemize}

\subsection{Primary dependencies}

The Compendium Manager uses the Click package (Ronacher and Lord 2024) for parsing user input and the SQLite library (Richard Hipp 2025) to build and maintain the database used for storing sample metadata, results, and project status information. The requests package (Reitz 2018) is used for HTTP calls. The example project-level workflow is automated using Snakemake (Köster and Rahmann 2012) to download samples using the SRA Toolkit (Sra-Tools: SRA Tools, n.d.) and process them using DADA2 (Callahan et al. 2016). ASV inference is performed using scikit-bio (Rideout et al. 2025).

\section{Conclusion}

The Compendium Manager software represents a proof of concept for an automation layer on top of Snakemake (or, in principle, any workflow management tool that can execute commands to run the Python application). Automation at this level is relevant is because of the rapidly expanding availability of sequencing data: Biological databases hold an overwhelming amount of data available for public use, and they have grown rapidly since early calls to gain novel insight by integrating many publicly available projects into unified analyses (Greene and Troyanskaya 2012). Those maintained by the National Center for Biotechnology Information (NCBI) in the United States, for example, contain billions of sequences describing DNA, RNA, proteins, and other biologically relevant compounds (Sayers et al. 2025), plus data on 32 million sequencing runs (Karsch-Mizrachi et al. 2025) occupying an exponentially growing number of petabytes of space (Katz et al. 2022). Some aspects of the current workflow require manual intervention---when some samples within a project are unavailable for download, or when a Snakemake pipeline exits with an error---but we believe the software's ongoing development as an open-source project will enable the compilation of integrative datasets at a novel scale.

\section*{References}
Abdill, Richard J., Samantha P. Graham, Vincent Rubinetti, Mansooreh Ahmadian, Parker Hicks, Ashwin Chetty, Daniel McDonald, et al. 2025. “Integration of 168,000 Samples Reveals Global Patterns of the Human Gut Microbiome.” \textit{Cell}. https://doi.org/10.1016/j.cell.2024.12.017.

Altschul, S. F., W. Gish, W. Miller, E. W. Myers, and D. J. Lipman. 1990. “Basic Local Alignment Search Tool.” \textit{Journal of Molecular Biology} 215 (3): 403--10.

Barrett, Tanya, Karen Clark, Robert Gevorgyan, Vyacheslav Gorelenkov, Eugene Gribov, Ilene Karsch-Mizrachi, Michael Kimelman, et al. 2012. “BioProject and BioSample Databases at NCBI: Facilitating Capture and Organization of Metadata.” \textit{Nucleic Acids Research} 40 (Database issue): D57--63.

Callahan, Benjamin J., Paul J. McMurdie, Michael J. Rosen, Andrew W. Han, Amy Jo A. Johnson, and Susan P. Holmes. 2016. “DADA2: High-Resolution Sample Inference from Illumina Amplicon Data.” \textit{Nature Methods} 13 (7): 581--83.

Clarke, Daniel J. B., Avi Ma’ayan, and Alexander Lachmann. 2020. archs4: ARCHS4 RNA-Seq Processing Scripts and Web Server Pages. Github. https://github.com/MaayanLab/archs4.

Dai, Die, Jiaying Zhu, Chuqing Sun, Min Li, Jinxin Liu, Sicheng Wu, Kang Ning, Li-Jie He, Xing-Ming Zhao, and Wei-Hua Chen. 2022. “GMrepo v2: A Curated Human Gut Microbiome Database with Special Focus on Disease Markers and Cross-Dataset Comparison.” \textit{Nucleic Acids Research} 50 (D1): D777--84.

Di Tommaso, Paolo, Maria Chatzou, Evan W. Floden, Pablo Prieto Barja, Emilio Palumbo, and Cedric Notredame. 2017. “Nextflow Enables Reproducible Computational Workflows.” \textit{Nature Biotechnology} 35 (4): 316--19.

Greene, Casey S., and Olga G. Troyanskaya. 2012. “Chapter 2: Data-Driven View of Disease Biology.” \textit{PLoS Computational Biology} 8 (12): e1002816.

Hawkins, Allegra G., Joshua A. Shapiro, Stephanie J. Spielman, David S. Mejia, Deepashree Venkatesh Prasad, Nozomi Ichihara, Arkadii Yakovets, et al. 2024. “The Single-Cell Pediatric Cancer Atlas: Data Portal and Open-Source Tools for Single-Cell Transcriptomics of Pediatric Tumors.” bioRxiv. https://doi.org/10.1101/2024.04.19.590243.

Home. n.d. Github. Accessed May 13, 2025. https://github.com/waldronlab/curatedMetagenomicDataCuration/wiki.

Karsch-Mizrachi, Ilene, Masanori Arita, Tony Burdett, Guy Cochrane, Yasukazu Nakamura, Kim D. Pruitt, Valerie A. Schneider, and On Behalf Of The International Nucleotide Sequence Database Collaboration. 2025. “The International Nucleotide Sequence Database Collaboration (INSDC): Enhancing Global Participation.” \textit{Nucleic Acids Research} 53 (D1): D62--66.

Katz, Kenneth, Oleg Shutov, Richard Lapoint, Michael Kimelman, J. Rodney Brister, and Christopher O’Sullivan. 2022. “The Sequence Read Archive: A Decade More of Explosive Growth.” \textit{Nucleic Acids Research} 50 (D1): D387--90.

Köster, Johannes, and Sven Rahmann. 2012. “Snakemake--a Scalable Bioinformatics Workflow Engine.” \textit{Bioinformatics} 28 (19): 2520--22.

Lachmann, Alexander, Denis Torre, Alexandra B. Keenan, Kathleen M. Jagodnik, Hoyjin J. Lee, Lily Wang, Moshe C. Silverstein, and Avi Ma’ayan. 2018. “Massive Mining of Publicly Available RNA-Seq Data from Human and Mouse.” \textit{Nature Communications} 9 (1): 1366.

Leipzig, Jeremy. 2017. “A Review of Bioinformatic Pipeline Frameworks.” \textit{Briefings in Bioinformatics} 18 (3): 530--36.

Mangul, Serghei, Thiago Mosqueiro, Richard J. Abdill, Dat Duong, Keith Mitchell, Varuni Sarwal, Brian Hill, et al. 2019. “Challenges and Recommendations to Improve the Installability and Archival Stability of Omics Computational Tools.” \textit{PLoS Biology} 17 (6): e3000333.

Marx, Vivien. 2013. “Biology: The Big Challenges of Big Data.” \textit{Nature} 498 (7453): 255--60.

Masera, Marco, Alessandro Leone, Johannes Köster, and Ivan Molineris. 2025. “Snakemaker: Seamlessly Transforming Ad-Hoc Analyses into Sustainable Snakemake Workflows with Generative AI.” arXiv [cs.SE]. https://doi.org/10.48550/ARXIV.2505.02841.

Ouzounis, Christos A. 2012. “Rise and Demise of Bioinformatics? Promise and Progress.” \textit{PLoS Computational Biology} 8 (4): e1002487.

Ouzounis, Christos A., and Alfonso Valencia. 2003. “Early Bioinformatics: The Birth of a Discipline--a Personal View.” \textit{Bioinformatics} (Oxford, England) 19 (17): 2176--90.

Pasolli, Edoardo, Lucas Schiffer, Paolo Manghi, Audrey Renson, Valerie Obenchain, Duy Tin Truong, Francesco Beghini, et al. 2017. “Accessible, Curated Metagenomic Data through ExperimentHub.” \textit{Nature Methods} 14 (11): 1023--24.

Pearson, W. R., and D. J. Lipman. 1988. “Improved Tools for Biological Sequence Comparison.” \textit{Proceedings of the National Academy of Sciences of the United States of America} 85 (8): 2444--48.

Pohl, Sebastian, Nourhan Elfaramawy, Artur Miling, Kedi Cao, Birte Kehr, and Matthias Weidlich. 2024. “How Do Users Design Scientific Workflows? The Case of Snakemake and Nextflow.” In \textit{Proceedings of the 36th International Conference on Scientific and Statistical Database Management}, 163:1--12. New York, NY, USA: ACM.

Reitz, Kenneth. 2018. “Requests-HTML.” Github.

Richard Hipp, D. 2025. SQLite. https://www.sqlite.org.

Rideout, Jai Ram, Greg Caporaso, Evan Bolyen, Daniel McDonald, Yoshiki Vázquez Baeza, Jorge Cañardo Alastuey, Anders Pitman, et al. 2025. Scikit-Bio/scikit-Bio: Scikit-Bio 0.6.3. Zenodo. https://doi.org/10.5281/ZENODO.14640761.

Ronacher, Armin, and David Lord. 2024. Click (version 8.1.8). https://click.palletsprojects.com/en/stable/.

Sayers, Eric W., Jeffrey Beck, Evan E. Bolton, J. Rodney Brister, Jessica Chan, Ryan Connor, Michael Feldgarden, et al. 2025. “Database Resources of the National Center for Biotechnology Information in 2025.” \textit{Nucleic Acids Research} 53 (D1): D20--29.

Sra-Tools: SRA Tools. n.d. Github. Accessed May 7, 2025. https://github.com/ncbi/sra-tools.

Valderrama, Benjamin, Paulina Calderon-Romero, Thomaz F. S. Bastiaanssen, Aonghus Lavelle, Gerard Clarke, and John F. Cryan. 2025. “The South American MicroBiome Archive (saMBA): Enriching the Healthy Microbiome Concept by Evaluating Uniqueness and Biodiversity of Neglected Populations.” bioRxiv. https://doi.org/10.1101/2025.04.03.647034.

Wilks, Christopher, Shijie C. Zheng, Feng Yong Chen, Rone Charles, Brad Solomon, Jonathan P. Ling, Eddie Luidy Imada, et al. 2021. “recount3: Summaries and Queries for Large-Scale RNA-Seq Expression and Splicing.” \textit{Genome Biology} 22 (1): 323.

Wirth, Wytamma, Simon Mutch, and Robert Turnbull. 2024. “Snk: A Snakemake CLI and Workflow Management System.” \textit{Journal of Open Source Software} 9 (103): 7410.

Wren, Jonathan D. 2016. “Bioinformatics Programs Are 31-Fold over-Represented among the Highest Impact Scientific Papers of the Past Two Decades.” \textit{Bioinformatics} (Oxford, England) 32 (17): 2686--91.

\end{document}